\documentclass{nature-wfig}
\usepackage{graphicx}

\def\arcsec{^{\prime\prime}}
\def\gtrsim{\mathrel{\hbox{\rlap{\hbox{\lower5pt\hbox{$\sim$}}}\hbox{$>$}}}}
\newcommand{\ee}[1]{\mbox{${} \times 10^{#1}$}}
\newcommand{\msun}{\mbox{M$_\odot$}}
\newcommand{\jj}[2]{\mbox{$J = #1\rightarrow#2$}}
\newcommand{\tgas}{\mbox{$T_{gas}$}}

\title{An Old Disk That Can Still Form a Planetary System}

\author{Edwin A. Bergin$^{1}$, L. Ilsedore Cleeves$^1$, Uma Gorti$^{2,3}$, Ke
Zhang$^{4}$, Geoffrey A. Blake$^{5}$, Joel D. Green$^6$, Sean M. Andrews$^7$,
Neal J. Evans II$^6$, Thomas Henning$^8$, Karin \"{O}berg$^7$, Klaus
Pontoppidan$^9$, Chunhua Qi$^7$, Colette Salyk$^{10}$, \& Ewine F. van
Dishoeck$^{11, 12}$}

\begin{document}

\maketitle

\begin{affiliations}
 \item Department of Astronomy, University of Michigan, 500 Church St., Ann
Arbor, MI 48109, USA
 \item SETI Institute, Mountain View, CA, USA
 \item NASA Ames Research Center, Moffett Field, CA, USA
 \item California Institute of Technology, Division of Physics, Mathematics \&
Astronomy, MS 150-21, Pasadena, CA 91125, USA
 \item California Institute of Technology, Division of Geological \& Planetary
Sciences, MS 150-21, Pasadena, CA 91125, USA
  \item Department of Astronomy, The University of Texas, 
   2515 Speedway, Stop C1402, Austin, TX 78712, USA
 \item Harvard-Smithsonian Center for Astrophysics, 60 Garden Street,
Cambridge, MA 02138, USA
  \item Max Planck Institute for Astronomy, K\"onigstuhl 17, 69117, Heidelberg,
Germany
   \item Space Telescope Science Institute, 3700 San Martin Drive, Baltimore,
MD 21218, USA
 \item National Optical Astronomy Observatory, 950 N. Cherry Ave., Tucson, AZ
85719, USA
 \item Max Planck Institut f\"{u}r Extraterrestrische Physik,
Giessenbachstrasse 1, 85748, Garching, Germany
 \item Leiden Observatory, Leiden University, PO Box 9513, 2300 RA, Leiden, The
Netherlands
\end{affiliations}

\begin{abstract}
From the masses of planets orbiting our Sun, and relative elemental
abundances, it is estimated that at birth
our Solar System required a minimum disk mass of
$\sim$0.01~M$_{\odot}$ within  $\sim$100 AU of the 
star\cite{kuiper_mmsn, kusaka_mmsn, w77, hayashi_mmsn}.
The main constituent, gaseous molecular hydrogen, does not emit from the disk mass 
reservoir\cite{carmona08}, so the most common
measure of the disk mass  is dust thermal emission and lines of gaseous carbon monoxide\cite{bs90}.  
Carbon monoxide emission generally probes the disk surface, while
the conversion from dust emission to gas mass requires knowledge of
the grain properties and gas-to-dust mass ratio, which
likely differ from their interstellar values\cite{h08, williams_araa}.  
Thus, mass estimates vary by orders of magnitude, as exemplified
by the relatively old (3--10 Myr) star TW Hya\cite{bn06,vs11}, 
with estimates ranging from 0.0005 
to 0.06 \msun\cite{weintraub89, calvet02,thi10, gorti11}.  Here we report
the detection the fundamental rotational transition 
of hydrogen deuteride, HD, 
toward TW Hya.
HD is a good tracer of disk gas because it follows the distribution of molecular hydrogen and its emission is sensitive to the total  mass.
The HD detection, combined with existing observations and detailed models, 
implies a disk mass 
$>$0.05~M$_{\odot}$, enough to form a planetary system like our own.
\end{abstract}


Commonly used  tracers of protoplanetary disk masses are thermal emission
from dust grains and rotational lines of carbon monoxide gas.
Both methods, however, rely on unconstrained assumptions.  The dust method
has to assume a dust opacity per gram of dust, and grain growth can change
this value dramatically\cite{natta_ppv}. The gas mass is then calculated by multiplying
the dust mass by the gas-to-dust ratio, usually assumed to be $\sim$100 from measurements of the interstellar medium\cite{draine07}. 
The gas mass thus rests on a large and uncertain correction factor.
The alternative is to use rotational CO lines as gas tracers, but these are
optically thick, and therefore trace the disk surface temperature, as opposed
to the midplane mass. 
The use of CO as a gas tracer then leads to large discrepancies between 
mass  estimates for different models of TW Hydrae 
(from $5 \times 10^{-4}$ M$_{\odot}$ to 0.06 M$_{\odot}$),
even though each matches a similar set of observations\cite{thi10, gorti11}.

Using the Herschel Space Observatory\cite{pilbratt10} PACS Spectrometer\cite{poglitsch10} we robustly detected (9$\sigma$)
the lowest rotational transition, \jj10, of hydrogen deuteride (HD)
in the closest ($D \sim 55$~pc)
and best studied circumstellar disk around TW Hydrae (Fig. 1). This star is
older (3-10 Million years\cite{hoff98,bn06,vs11}) than most stars with gas-rich
circumstellar disks\cite{williams_araa}.
The abundance of deuterium atoms relative to hydrogen is well
characterized via atomic electronic transitions to be 
$x_{\rm D} = 1.5 \pm 0.1 \times 10^{-5}$ 
in objects that reside within $\sim$ 100 pc of the Sun\cite{linsky98}.   Placing 
atoms into H$_2$ and HD, which is appropriate for much of the disk mass,  
provides an HD abundance relative to H$_2$ of $x_{\rm HD} = 3.0 \times 10^{-5}$.  
We combine the HD data with existing molecular observations to set new
constraints on the disk mass within 100 AU  -- the most fundamental 
quantity that determines whether planets can form. 
The disk mass also governs the primary mode of giant
planet formation, either through core accretion or gravitational
instability\cite{lissauer07}.
In this context we do not know whether our own Solar System formed within a typical disk, as nearly half of current disk mass estimates fall below the minimum mass solar nebula\cite{williams_araa}. Our current census of extra-solar planetary systems furthermore suggests that even larger disk masses are necessary to form many of the planetary systems seen\cite{gr10, mordasini12}.


With smaller rotational energy spacings  and a weak electric dipole moment,  HD
\jj10\ is a million times more emissive than H$_2$ for a given gas mass at a gas temperature of 
$\tgas = 20$ K.
The HD line flux ($F_l$) sets a lower limit to the H$_2$ gas mass at distance $D$  
(see Supplementary Information):

\small
\begin{equation}
M_{gas\;disk} > 5.2 \times 10^{-5} \left(\frac{F_l}{6.3 \times
10^{-18}\;{\rm W\;m^{-2}}}\right)\left(\frac{3 \times 10^{-5}}{x_{\rm
HD}}\right)\left(\frac{D}{55\;{\rm pc}}\right)^2\; \exp\left(\frac{128.5\;{\rm
K}}{T_{gas}}\right) \;{\rm M_{\odot}}.
\end{equation}

\normalsize
\noindent  If HD is optically thick or D is hidden in other molecules
such as PAHs or molecular ices, the conversion from deuterium mass to hydrogen 
mass will be higher and thus the mass will be larger, hence the lower limit.  
The strong temperature dependence arises from the fractional
population of the J=1 state which has a value of $f_{J=1} \sim
3\exp(-128.5\;{\rm K}/T_{gas})$ for $T_{gas} < 50$~K in thermal equilibrium.
Due to the low fractional population in the J=1 state, HD does not emit
appreciably from gas with $T \sim 10 - 15$~K, the estimated temperature in the
outer disk mass reservoir (R $\gtrsim$ 20--40 AU).
The HD mass derived from Eqn. (1) provides an estimate of the mass in  warm gas
gas, and is therefore a lower limit to the total mass
within 100 AU.  

The only factor in Eqn. (1) that could lower the mass estimate is a 
higher \tgas.  The upper limit on the \jj21\ transition of HD (Fig. 1)
implies 
$\tgas < 80$~K in the emitting region. This \tgas\ estimate yields 
$M_{gas\;disk} > 2.2\ee{-4}$ \msun, but \tgas\ is unlikely to be this high for the bulk of the disk.
CO rotational transitions are optically thick and the level populations
are in equilibrium with \tgas, so they provide a measure of \tgas.
ALMA observations of CO \jj32\ emission
in a 1.7$''$ $\times 1.5''$ beam
(radius $\sim 43$~ AU) 
(see Supplementary Information and Supplementary Figure 1)
yield an average $\tgas = 29.7$~K within 43 AU, and 
$M_{gas\;disk} > 3.9 \times 10^{-3}$~M$_{\odot}$.
This value is still likely to be too low because the
optically thick CO emission presumably probes material closer to the surface 
than HD, and this gas will be warmer than the HD line emitting region.
Thus essentially all correction factors would push the
mass higher than this conservative limit, which already rules out a portion
of the low end of previous mass determinations. 

To determine the mass more accurately, 
we turn to detailed models that incorporate explicit
gas thermal physics providing for substantial radial and vertical thermal
structure.
Both published models of the TW Hya disk reproduce a range of
gas phase emission lines,
but in one case with M$_{gas-disk} =$ 0.06 M$_{\odot}$\cite{gorti11} and in the
other with M$_{gas-disk} =$ 0.003 M$_{\odot}$\cite{thi10} (see Supplementary
Information and Supplementary Table S1).  These models were both placed into
detailed radiation transfer simulations.
The results from this calculation and the adopted physical structure are given
in Fig.~2 for the model with M$_{gas-disk} =$ 0.06 M$_{\odot}$.
Fig.~2c shows the cumulative flux as a function of radius for the higher mass
model; over 80\% of the emission is predicted to arise from gas inside of 80
AU.    Furthermore Fig.~2d provides a calculation of the HD emissive mass as a
function of gas temperature.  This calculation suggests gas with temperatures of
30--50 K is responsible for the majority of the HD emission.

The model with M$_{gas-disk} =$ 0.003 M$_{\odot}$ predicts an HD line flux of
$F_l = 3.8 \times 10^{-19}$ W m$^{-2}$,
more than an order of magnitude below the detected level.
For this model to reach the observed flux the disk mass must be
increased by a factor of 20, ruling out this low disk mass.
The M$_{gas-disk} =$ 0.06 M$_{\odot}$ model predicts 
$F_l = 3.1 \times 10^{-18}$ W m$^{-2}$, still a factor of two below 
the observed value.  Even the ``high" mass estimate is too low.
Based on this model we estimate the disk gas mass within 80 AU, where the majority of HD emissions arise, is
$0.056$ M$_{\odot}$.
Both of these models match other observations as the low disk mass model matches CO and $^{13}$CO \jj32, while the higher mass model reproduces CO \jj21, \jj32,  \jj65.  Both compare emission predictions to other species as well.   However, they differ by a factor of 10 in predicting the HD 
emission.
This difference shows the value of HD in constraining masses.

The age of TW Hya is uncertain.  The canonical age of the cluster is 10$_{-7}^{+10}$ Million years\cite{bn06}.  
However there could be an age spread in cluster members and ages estimated for TW Hya itself
range from 3--10 Myr\cite{hoff98,vs11}.
Even at the low end of this range TW Hya is older than the half-life of gaseous disks, inferred to be about
2 Myr\cite{williams_araa}.    In the case of the TW Hya association there is also little evidence for an associated molecular cloud\cite{tachihara09}, which is an additional indicator that this system is {\em relatively} older than most gas-rich disks.
 The lifetime of the gaseous disk is important as it sets the available timeframe for the formation of gas giants equivalent to Jupiter or Saturn.
Based on our analysis, TW Hya contains
a massive gas disk ($\gtrsim 0.06$~M$_{\odot}$) that is several times the minimum mass required to make the
planets in our solar system.   Thus, this ``old'' disk can still 
form a planetary system like our own.  

The recent detection of cold water vapor from TW Hya found indirect evidence for a large water ice reservoir ($\sim$ several thousand Earth oceans) assuming a disk mass of 0.02 M$_{\odot}$\cite{hoger11a}.  Our higher mass estimate implies a larger water ice reservoir, perhaps increased by a factor of two.
The mass estimate in this one system lies on the upper end of
previous mass measurements\cite{williams_araa}, hinting that other disk masses
are underestimated. The main uncertainty in the masses derived here is
the gas temperature structure of the disk.
 Looking forward, observations of optically thick
molecular lines, particularly CO, can be used to trace the {\em thermal structure of
 gas} in the disk.   Observations of rarer CO isotopologues then provide
constraints on the temperature in deeper layers\cite{ddg03}.   With 
the Atacama Large Millimeter Array,  we will readily
resolve multiple gas temperature tracers inside 80 AU where HD strongly
emits.
When used in tandem with HD we will be able to derive the gas mass with much
greater accuracy (our simulations suggest to within a factor of 2--3).
Moreover, additional HD detections could be provided by Herschel and with higher spectral
resolution by SOFIA GREAT under favorable atmospheric conditions. 
These data could be used alongside emission from species such as
C$^{18}$O, C$^{17}$O, or the dust to calibrate these more widely
available probes to determine the disk gas mass.   
Thus, with the use of HD to
complement other observations  and constrain models
we may
finally place useful constraints on one of the most important quantities that
governs the process of planetary birth.


\begin{figure}
\centering
\includegraphics[height=14cm]{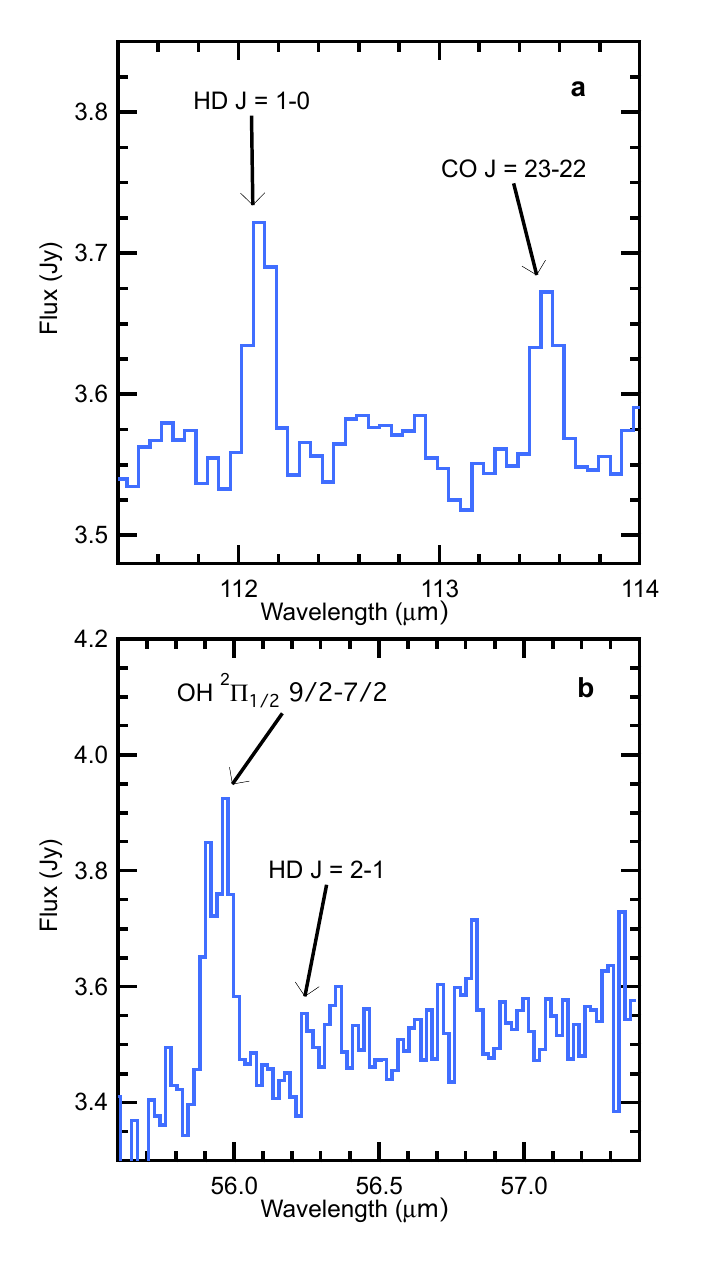}
\caption{{\bf Herschel detection of Hydrogen Deuteride  in the TW Hya
protoplanetary disk.} \small
  (a) The
 fundamental \jj10\ line of HD lies at $\sim$112$\mu$m line.  On 20 November 2011 it was detected towards the TW Hya disk at
the level of 9$\sigma$.  The total integrated flux is  $6.3 \pm 0.7 \times
10^{-18}$ W m$^{-2}$.   We also report a detection of the warm disk atmosphere
in CO $J = 23 \rightarrow 22$  with a total integrated flux of $4.4 \pm 0.7 \times
10^{-18}$ W m$^{-2}$.  The \jj10\ line of HD was previously detected in a
warm gas cloud exposed to radiation from nearby stars\cite{wright99} by the
Infrared Space Observatory.  Other transitions have also been detected in
shocked regions associated with supernova and outflows from massive
stars\cite{bertoldi99, neufeld_hd}.
 (b) Simultaneous observations of HD \jj21\ are shown.  For HD \jj21\ we find a
detection limit of $< 8.0$ $\times$ $10^{-18}$ W m$^{-2}$ (3$\sigma$).   We also
report a detection of the OH $^2\Pi_{\frac{1}{2}}$ $\frac{9}{2} \rightarrow \frac{7}{2}$
doublet near 55.94 $\mu$m with an integrated flux of $4.93 \pm 0.27$ $\times
10^{-17}$  W m$^{-2}$.
 The spectra include the observed thermal dust continuum of $\sim 3.55$~Jy at
both wavelengths. }
\end{figure}

\begin{figure}
\centering
\includegraphics[height=12cm]{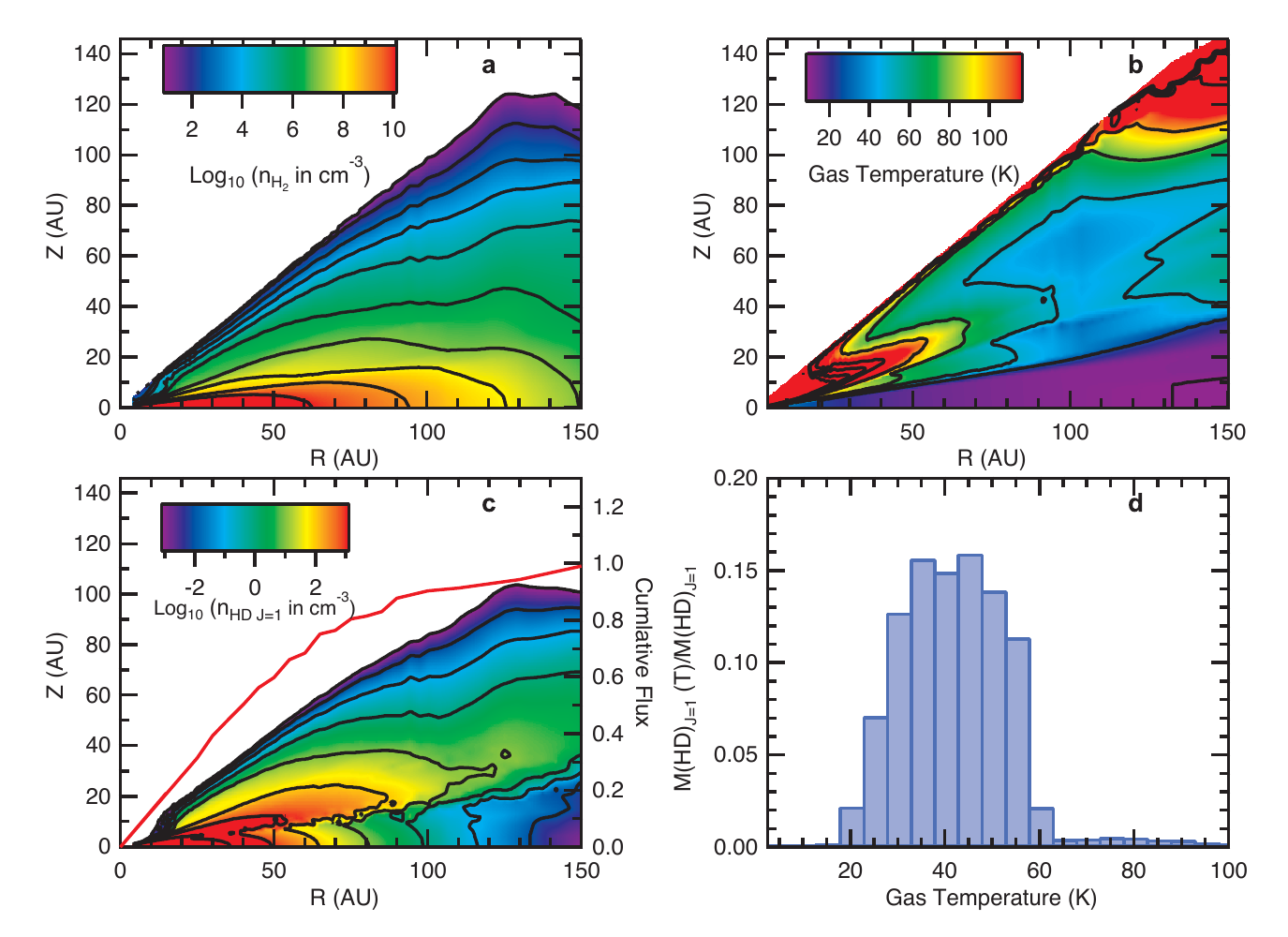}
\caption{{\bf Model of the physical structure and HD emission in the TW Hya
circumstellar disk.} \small
(a) Radial and vertical distribution of the H$_2$ volume density, $n_{\rm H_2}$, calculated in a
model disk with mass 0.06 M$_{\odot}$\cite{gorti11}.
Contours start from the top at Log$_{10}$ [$n_{\rm H_2}$/cm$^{-3}$] = 1.0 and step
downwards in units of factors
of ten.  (b) The gas temperature structure as derived by the thermochemical
model\cite{gorti11}.  Contours are 10, 25, 50, 75, 100, 150, 200, 250, and 300 K.
(c) Radial and vertical distribution of the HD J = 1 volume density, $n_{\rm HD\; J=1}$, 
predicted in a model disk with the gas density and temperature
structure as given in panels (a) and (b), with an HD abundance relative to
H$_2$ of $3.0 \times 10^{-5}$.
Contours start at the top with Log$_{10}$ [$n_{\rm HD\; J=1}$/cm$^{-3}$] = $-3$ and
are stepped in factors of 10.  The red line on the plot shows the cumulative
flux contribution as a function of radius in terms of fractions of the overall
predicted flux of $3.1 \times 10^{-18}$ W m$^{-2}$.
To predict the HD line emission we calculate the solution of the equations of
statistical equilibrium including the effects of line and dust opacity using
the LIME code\cite{lime}.    (d) Fraction of the HD emission arising from gas
with different temperatures, computed as function of the mass of HD excited
to the J=1 state in gas at temperatures binned in units of 5 K and normalized
to the total mass of HD in J=1.   In particular, $\int n_{\rm HD\; J=1}2\pi Rdrdz$
is computed successively in gas temperature bins of 5 K and then normalized to
the total mass of HD in the J $=$ 1 state.
}
\end{figure}


\newcommand{\nat}{{ Nature }}
\newcommand{\aap}{{Astron. \& Astrophys. }}
\newcommand{\aj}{{ Astron.~J. }}
\newcommand{\apj}{{ Astrophys.~J. }}
\newcommand{\araa}{{Ann. Rev. Astron. Astrophys. }}
\newcommand{\apjl}{{Astrophys.~J.~Letters }}
\newcommand{\apjs}{{Astrophys.~J.~Suppl. }}
\newcommand{\apss}{{Astrophys.~Space~Sci. }}
\newcommand{\icarus}{{Icarus }}
\newcommand{\mnras}{{MNRAS }}
\newcommand{\pasp}{{ Pub. Astron. Soc. Pacific }}
\newcommand{\pasj}{{ Pub. Astron. Soc. Japan }}
\newcommand{\ssr}{{Space Sci. Rev.}}
\newcommand{\planss}{{Plan. Space Sci. }}
\newcommand{\physrep}{{ Phys. Rep.}}
\newcommand{\bain}{{Bull.~Astron.~Inst.~Netherlands }}
\newcommand{\pra}{{Phys.~Rev.~A}}%

\bibliography{/Users/ebergin/tex/bib/ted}


\begin{addendum}
 \item Herschel is an ESA space observatory with science instruments provided
by European-led Principal Investigator consortia and with important
participation from NASA.
Support for this work was provided by NASA through an award issued by
JPL/Caltech and by the National Science Foundation under grant 1008800.  This
paper makes use of the following ALMA data: ADS/JAO.ALMA\#2011.0.00001.SV. ALMA
is a partnership of ESO (representing its member states), NSF (USA) and NINS
(Japan), together with NRC (Canada) and NSC and ASIAA (Taiwan), in cooperation
with the Republic of Chile. The Joint ALMA Observatory is operated by ESO,
AUI/NRAO and NAOJ.

 \item[Author Contributions]  EAB, IC, UG, and KZ performed the detailed
calculations used in the analysis.  JG reduced the Herschel data.  SA provided
detailed disk physical models and UG provided thermochemical models, both
developed specifically for TW Hya.   EAB wrote the manuscript with revisions by NJE.  All authors
were participants in the discussion of results, determination of the
conclusions, and revision of the manuscript.

 \item[Competing Interests] The authors declare that they have no
competing financial interests.
 \item[Correspondence] Correspondence and requests for materials
should be addressed to Edwin Bergin~(email: ebergin@umich.edu).







\noindent {\bf Supplementary Information}
 
\hspace{-0.2in} {\bf 1.   Herschel Observations}

TW Hya ($\alpha(2000)$ = 11$^h$01$^m$51.91$^s$; $\delta(2000)$ =
$-34^{\circ}$43$'$17.0$''$)
 was observed as part of an open time Herschel Program (OT1\_ebergin\_4) using
the PACS$^{18}$ instrument on November 20, 2011.  The  PACS Range
scan chop-nod mode was used.  The background emission from the the 
telescope and sky was
subtracted using two nod positions 1.5$'$ distant from the source in each
direction.  The data cover a small spectral range (111.2-114.2 $\mu$m and
55.6-57.1 $\mu$m), with high sampling density (for a total of 81 steps).   The
first range includes HD \jj10\ and the second was designed to detect HD 
\jj21.
We used 12 repetitions to increase sensitivity for a deep scan, for a
total integration time of 25124 seconds (36 repetitions) on TW Hya.   The
predicted line RMS was 0.62 $\times$ 10$^{-18}$ W m$^{-2}$ 
and 2.65 $\times$ 10$^{-18}$ W m$^{-2}$ at 112 and 56 $\mu$m, respectively.

The data were reduced using the Herschel Interactive Processing Environment
(HIPE)\cite{hipe} v8.1 pipeline (calibration set 32). PACS is a 5$\times$5
integral field unit spectrometer with a pixel size of 
9.4$\arcsec$ $\times$9.4$\arcsec$; 
the source showed no sign of extended emission beyond the point
spread function and was centered within 0.2 pixels (within 2$\arcsec$) of the
centerpoint in each case.  We used the standard ``calibration block'' script
to reduce the data for optimal signal-to-noise utilizing only the 
central pixel  of the array, 
and then scaled this value to the spectra extracted from the
central 3$\times$3 pixels.  We compared this to the PSF-corrected output from
the pipeline, and the flux levels matched to within 10\%.
The PSF-corrected flux was 10\% {\it higher} than the 3$\times$3 extraction,
which indicates an overcorrection and no sign of extended emission.  Thus, we
increased the flux measured from the central-spaxel-only spectrum by 10\% to
match that of the 3$\times$3 extraction, yielding line fluxes, 
$F_l =  (4.4 \pm 0.7) \times 10^{-18}$ W m$^{-2}$ for 
CO \jj{23}{22}\ (centered at 113.509 $\mu$m indicating a possible blend 
between CO at 113.46 $\mu$m and H$_2$O at 113.53 $\mu$m), 
and $F_l = (6.3 \pm 0.7) \times 10^{-18}$ W m$^{-2}$ for HD \jj10\
(centered at 112.086 $\mu$m).
The line flux was obtained using a Gaussian fit.  
The continuum was determined by a
first-order polynomial simultaneous fit in a  fairly tight region around the
line, avoiding the CO transition.
The HD \jj21\ transition was not detected with a 3$\sigma$ upper limit of
$F_l < 8.0 \times 10^{-18}$ W m$^{-2}$.  The lines are not resolved with
$\lambda/\Delta \lambda \sim 1000$ at 112 $\mu$m and $\sim 1500$ at 56 $\mu$m.
The absolute uncertainty on flux calibration is potentially
larger, with an important factor being the overall pointing and
presence/absence of extended flux.  Because TW Hya is a point source, the
observations are well centered, and we included a PSF correction,
the uncertainty in $F_l$ is limited to 10-20\%, negligible compared
to other uncertainties.

 From the CDMS database\cite{muller05}, the \jj10\ line at 2674.986 GHz has
$E_{J=1} = 128.5$~K with $A_{10} = 5.44 \times 10^{-8}$ s$^{-1}$.  The \jj21\
line at 5331.561 GHz has $E_{J=2} = 384.58$~K with $A_{21} = 5.16 \times
10^{-7}$ s$^{-1}$.
 The A-coefficients have been calculated using the dipole moment of  8.56
$\times 10^{-4}$~D\cite{pk08}.


\noindent {\bf 2. Simple Estimate of the Gas Mass}

The mass implied by a line flux of optically thin emission from an unresolved 
source can be derived in two steps. The total number of HD molecules
($\mathcal{N_{\rm HD}}$) is related to the line flux by this relation, assuming that the beam
encompasses the source:

 \begin{equation}
 F_l = \frac{\mathcal{N_{\rm HD}}  A_{10} h \nu f_{u}}{4 \pi D^2}.
 \end{equation}

  \noindent In this expression $f_u = 3.0 * exp(-128.5\;{\rm K}/T)/Q(T)$ is the
fractional of HD molecules in $J = 1$, $D$ is the distance, and $\nu$ is the
frequency.
Converting to mass, and assuming all is in H$_2$, $M_{gas\;disk} = 2.37 * m_{\rm H} \mathcal{N_{\rm HD}}/
x_{\rm HD}$, where $x_{\rm HD}$ is the abundance of HD relative to H$_2$,
$m_{\rm H}$ is the mass of a hydrogen atom, and 2.37 is the mean 
molecular weight per particle, including helium and heavy elements\cite{kauffmann08}.
This gives

  \begin{equation}
  M_{gas\;disk} = \frac{2.37 m_H 4 \pi D^2 F_l}{A_{10} h \nu x_{\rm
HD} f_u}.
  \end{equation}

 The partition function, $Q(T)$, is near unity below $\sim 50$
K\cite{muller05}.  Inserting values of physical constants yields:

\begin{equation}
M_{gas\;disk} > 5.21 \times 10^{-5} \left(\frac{F_l}{6.3 \times
10^{-18}\;{\rm W\;m^{-2}}}\right)\left(\frac{3 \times 10^{-5}}{x_{\rm
HD}}\right)\left(\frac{D}{55\;{\rm pc}}\right)^2\; \exp\left(\frac{128.5\;{\rm
K}}{T_{gas}}\right) \;{\rm M_{\odot}}.
\end{equation}

\noindent 
This  estimate represents a lower limit because the HD emission may
not trace all the mass in the disk.

  \noindent {\bf 3. CO Emission and $T_{gas}$ from TW Hya}

In the main text we used the resolved ALMA Science Verification CO \jj32\ data to
set a limit on the gas temperature.
In Supplementary Figure 1 we show the integrated emission map with a beam size
1.7$''$ $\times 1.5''$, which
corresponds to a radius of 47 $\times$ 41 AU.   The map demonstrates that
the CO emission is resolved, so we assume unity filling factor.
Based on the central spectrum the observed peak radiation temperature
is $T_R = 22.2\pm 0.1$~K (we note that this data has a calibration uncertainty of 10\%). 
$T_R$ is linearly related to the intensity, so 
a correction for the difference between the Planck function and the 
Rayleigh-Jeans approximation results in an
average gas kinetic temperature of 
$T_{gas} = 29.7$ K in the beam, assuming optically thick
and thermalized emission.
Observations of CO \jj65\ confirm this value.
The observed CO \jj65\ peak intensity is T$_{R} = 16.9 \pm 3.2$ K\cite{qi06}, 
which corresponds to a gas temperature of 30.6 K.  

Assuming $T_{gas} = 30$ K yields a minimum disk mass of 3.9\ee{-3} 
\msun.

   \noindent {\bf 4. HD Emission Line Models}


For more detailed models we use two predicted physical structures (Gorti et al.$^{14}$
and the Thi et al.$^{13}$ TW Hya models) derived from thermochemical calculations that
match a range of gas phase emission lines.  The Gorti et al. TW Hya model has a disk
gas mass of 0.06 M$_{\odot}$, and we used direct output from the model.    For
the Thi et al. TW Hya model we use a reproduction.  Specifically, to reproduce this model we
adopt the physical parameters for the dust structure  as given in their$^{13}$
Table 2.    The dust optical constants are as prescribed\cite{dorschner95} with
a dust grain size distribution from $3 \times 10^{-2}$ $\mu$m to 1 mm and a
power law index of 3.4.  These inputs were placed into RADMC\cite{dd04} where
we verified that the model reproduced the observed spectral energy
distribution and the dust thermal structure given in their Fig.  A.2. 
For this calculation the total disk gas mass is 0.003~M$_{\odot}$ 
and we verified that our gas density distribution matched that
provided in their Figure A.1.   We do not model the H-H$_2$ transition; this is appropriate for
HD which will emit below this transition region.    Finally we used the gas temperature
distribution given in their Fig. A.3 as input to our LIME calculations for HD.  As a final check
we computed the predicted emission of this reproduction for CO and $^{13}$CO \jj32\ to observed values\cite{vanz_etal01}
and found them to be in agreement to within a factor of two.

To predict the HD line emission from the physical models we calculate the
solution of the equations of statistical equilibrium including the effects of
line and dust opacity using the LIME code$^{30}$.    Although the HD line
emission is fairly close to LTE, we adopt the collision rate coefficients of HD
with H$_2$\cite{flower00}.  The collision rates are insensitive to
the rotational state of the H$_2$ collision partner\cite{flower99}.
 The disk dust optical depth is determined by the
dust opacity coefficient $\kappa_{\lambda}$ and the dust mass distribution
within the disk.  Typical dust mixtures\cite{dorschner95, pollack94}
 suggest $\kappa_{112\;\mu m}$ $\sim$ 30 cm$^2$ g$^{-1}$.  The excitation model
 explicitly explores pumping by continuum radiation, which is found to be 
 negligible when compared to the effects of the dust optical depth.

 Table S1 provides the predictions from the specific thermochemical model
including an emission calculation with  dust opacity at 112 $\mu$m (the
realistic case) and without dust opacity.  The Thi et al. TW Hya model with the lowest
mass predicts an HD emission line that is too weak.  To match observations
within the framework of the Thi et al. TW Hya model we require an increase in mass by
a factor of 20. We stress that this is approximate as scaling the mass will
change the thermal solution (gas temperature, chemistry).  Thus, this factor
only illustrates the mass dependency and is not definitive.
   The thermochemical model that comes closest to the observed flux is the
Gorti et al. TW Hya 0.06 M$_{\odot}$ model -- which is still
about a factor of two smaller than the observed HD emission.

\end{addendum}

%
%

\begin{center}
\begin{tabular}{lrrr}
\multicolumn{4}{c}{\bf Supplementary Table S1: Specific Model Predictions}
\\\hline
\multicolumn{1}{c}{\bf Disk Model} &
\multicolumn{1}{c}{\bf Gas Mass} &
\multicolumn{1}{c}{\bf HD \jj10$^a$} &
\multicolumn{1}{c}{\bf HD \jj21} \\
\multicolumn{1}{c}{} &
\multicolumn{1}{c}{(M$_{\odot}$)} &
\multicolumn{1}{c}{(W m$^{-2}$)} &
\multicolumn{1}{c}{(W m$^{-2}$)} \\\hline\hline
Thi  et al.  TW Hya & 0.003  & 3.8 $\times 10^{-19}$ & 1.4 $\times 10^{-19}$ \\
Gorti et al. TW Hya & 0.06 & 3.1 $\times 10^{-18}$ & 3.3 $\times 10^{-18}$
\\\hline\hline
Observations & & 6.3$\pm$0.7 $\times 10^{-18}$  & $<$8.0 $\times 10^{-18}$ \\
\hline
\multicolumn{4}{l}{\footnotesize $^a$Fluxes without dust optical depth are 7.4
$\times 10^{-19}$ W m$^{-2}$ (Thi et al. 0.003 M$_{\odot}$),} \\
\multicolumn{4}{l}{\footnotesize \hspace{0.05in}4.2 $\times 10^{-18}$ W
m$^{-2}$ (Gorti et al. 0.06 M$_{\odot}$)}\\
\end{tabular}
\end{center}

\renewcommand{\figurename}{\bf Supplementary Figure}
\begin{figure}
\centering
\includegraphics[height=11cm]{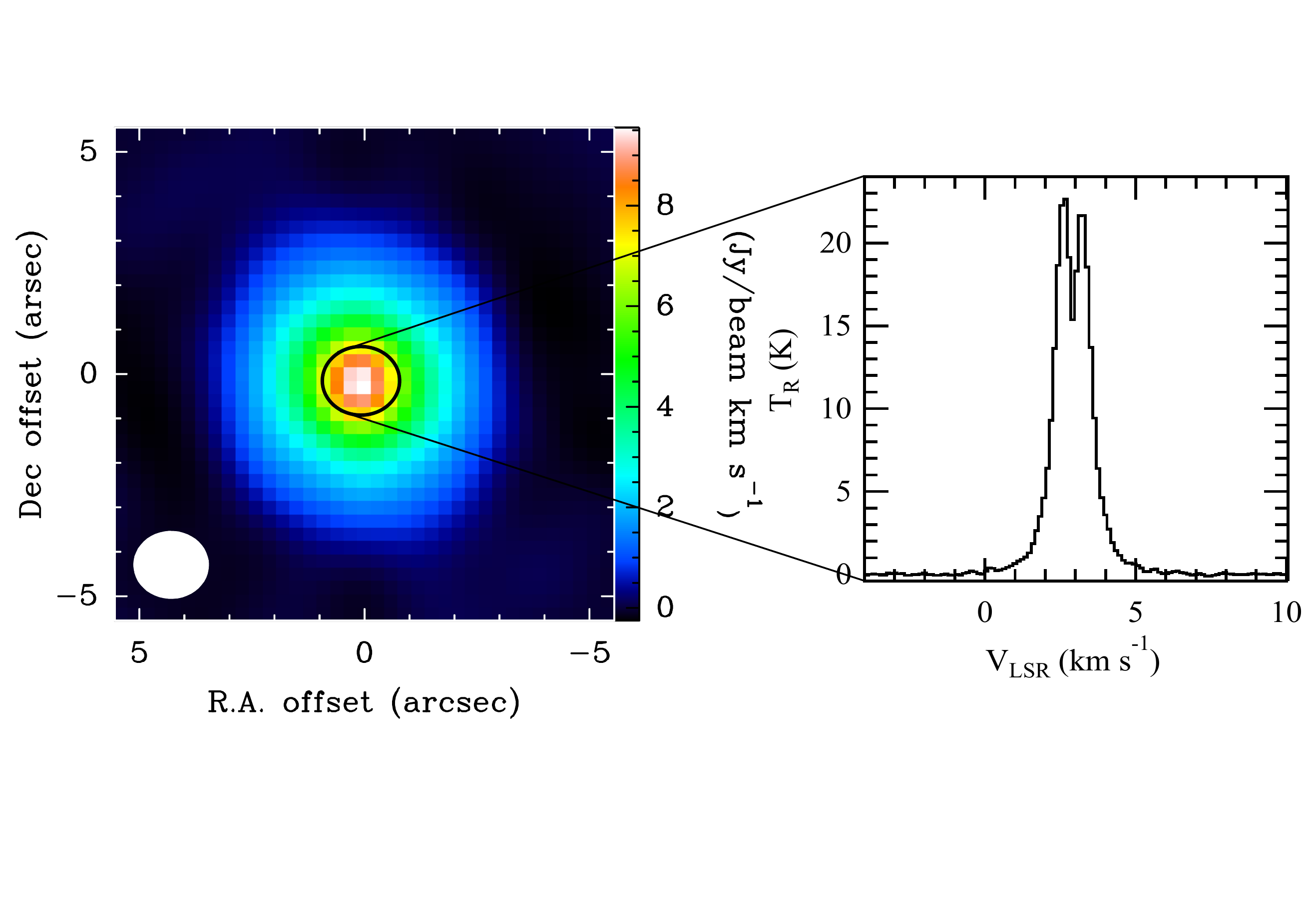}
\caption{{\bf ALMA Science Verification Observations of CO \jj32\ in TW
Hydra.}
(Left)  Map of CO \jj32\ integrated emission with intensity scale given on
the left.   The beam size of this observation is 1.7$'' \times 1.5''$ and is shown in the figure.  (Right)
Blow-up of the central spectrum. }
\end{figure}

\end{document}